\documentclass[aps,prc,reprint,superscriptaddress,nofootinbib]{revtex4-1}
\usepackage{graphicx}
\usepackage{multirow}
\usepackage{comment}
\usepackage{color}
\usepackage{ulem}

\usepackage{amssymb}
\usepackage{amsmath}
\usepackage{lineno}

\def\nuc#1#2{\relax\ifmmode{}^{#1}{\protect\text{#2}}\else${}^{#1}$#2\fi}

\newcommand{\vecr}{{\vec r}}
\newcommand{\vecR}{{\vec R}}
\newcommand{\vecK}{{\vec K}}

\newcommand{\half}{\frac{1}{2}}
\newcommand{\fhalf}{\frac{5}{2}}
\newcommand{\thalf}{\frac{3}{2}}

\newcommand{\be}{\begin{eqnarray}}
\newcommand{\ee}{\end{eqnarray}}

\newcommand{\bwt}{\begin{widetext}}
\newcommand{\ewt}{\end{widetext}}
\newcommand{\threej}[6]{\begin{pmatrix}#1&#2&#3\\#4&#5&#6\end{pmatrix}}

\begin{document}
%\begin{CJK*}{GB}{song}
%\preprint{}


%\title{Simultaneous inclusion of target excitations and deuteron breakup in $(d,p)$ transfer reactions. }  
\title{Influence of target deformation and deuteron breakup in $(d,p)$ transfer reactions}


\author{M. Gomez-Ramos}
\email[]{mgomez40@us.es}
%\thanks{}
\affiliation{Departamento de FAMN, Universidad de Sevilla, 
Apartado 1065, 41080 Sevilla, Spain.}

\author{A. M. Moro}
\email[]{moro@us.es}
%\thanks{}
\affiliation{Departamento de FAMN, Universidad de Sevilla, 
Apartado 1065, 41080 Sevilla, Spain.}


\begin{abstract}
\begin{description}
\item[Background] The effect of core excitations in transfer reactions of the form $A(d,p)B$ has been reexamined by some recent works, using the Faddeev/AGS reaction formalism. The effect was found to affect significantly the calculated cross sections and  
% by comparing calculations with and without including these effects, using as reaction model a momentum-space formulation of the  Faddeev equations. 
to depend strongly and non-linearly on the incident deuteron energy.

\item[Purpose] Our goal is to investigate  these effects within a coupled-channels formulation of the scattering problem which, in addition of being computationally less demanding than the Faddeev counterpart, may help shed some light into the physical interpretation of the cited effects. 

\item[Method] We use an extended version of the continuum-discretized coupled-channels (CDCC) method with explicit inclusion of target excitations within a  coupled-channels Born approximation (CDCC-BA) formulation of the transfer transition amplitude. We compare the calculated transfer cross sections with those obtained with an analogous calculation omitting the effect of target excitation. We consider also an  adiabatic coupled channels (ACC) method.  Our working example is the $^{10}$Be($d$,$p$)$^{11}$Be reaction. 

\item[Results]  We find that the two considered methods (CDCC-BA and ACC) reproduce fairly well the reported energy dependence of the core excitation effect. The main deviation from the pure three-body model calculation (i.e., omitting core excitations) is found to mostly originate from the destructive interference of the direct one-step transfer, and the two-step transfer following target excitation.
% Moreover, the sign of this effect is linked to the relative phase of the different components of the transferred particle in the residual nucleus.  

\item[Conclusions]  The proposed method, namely, the combination of the CDCC method and the CCBA formalism, provides a useful and accurate tool to analyse transfer reactions including explicitly, when needed, the effect  of target excitations and projectile breakup.  The method could be useful for other transfer reactions induced by weakly-bound projectiles, including halo nuclei.    
\end{description} 
\end{abstract}

\date{\today}

\maketitle

% \linenumbers
\section{Introduction \label{sec:intro}}
Transfer reactions have been used over the years as spectroscopic tools for extracting
 spin-parity assignments for nuclear states,  spectroscopic strengths of single-particle configurations, and  asymptotic normalization coefficients characterizing the tail of overlap functions. Many analyses of transfer reactions resort to the distorted-wave Born approximation (DWBA) method, which  can be regarded as the leading term of the Born expansion in terms of a transition potential, and assumes that the reaction is dominated by the elastic channel.  The effect of non-elastic channels is assumed to be effectively taken into account by the entrance channel
%by representing the entrance channel projectile--target relative motion
  optical model potential describing elastic scattering. Furthermore, very often, this optical potential, which would be angular-momentum dependent and non-local, is represented through a simple potential parametrization, for instance, of Woods-Saxon form, containing central and possibly, spin-orbit terms. This approach was early recognized to have severe limitations. First, it is not obvious that the effect of non-elastic channels on the calculated transfer cross sections is properly taken into account by the entrance channel optical potential. Further, it is well known that elastic scattering between heavy ions is mostly sensitive to the nuclear surface, whereas the transfer process is sensitive to small separations between the transferred particle and the respective cores to which it is initially or finally bound. Thus, the approximated three-body wave function used in DWBA, consisting of a product of the elastic scattering optical potential wave function times the projectile and target ground states wave functions, is not necessarily accurate for the transfer process. 

To overcome these shortcomings, appropriate extensions and alternative models have been proposed and applied. These extensions tend to emphasize specific aspects of the reaction dynamics. For example, when collective excitations are relevant, these can be included by means of a coupled-channels description of the entrance and/or exit channels. This is the {\it coupled-channels Born approximation} (CCBA) \citep{Asc69,Asc74,Gle71,Mac71}. For reactions induced by weakly-bound nuclei, such as deuterons, breakup channels are known to be important and must therefore be taken into account. This has been done in a number of ways. One of the most widespread approaches is the adiabatic distorted wave approximation (ADWA) method first proposed by Johnson and Soper \cite{Ron70} and later improved by these and other authors \cite{Har70,Ron74}.  The ADWA transition amplitude is formally identical to that appearing in DWBA, allowing its implementation in standard DWBA codes. The adiabatic model frequently provides significant improvements over DWBA for $A(d,p)B$ reactions. A more elaborated way of including the effect of the breakup channels is by means of a continuum-discretized coupled-channels (CDCC) expansion of the $d+A$ three-body wave function \cite{Aus87,Mor09,Kee04,Chau11}. 

For $(d,p)$ reactions on deformed targets, one may anticipate that both projectile breakup and target excitation can play a role and may require their explicit inclusion. This is not possible in the standard formulations of the aforementioned CCBA and ADWA methods, which tend to emphasize one of the two mechanisms. An exception is provided by the  Faddeev formalism, which has been recently extended to include collective excitations of the fragments (globally referred to as {\it core excitations}) \cite{Del13,Del15}. The core excitation effect was found to affect the calculated transfer cross sections beyond the expected scaling of the cross section due to the  corresponding spectroscopic factors. The problem has been further investigated in a recent work \cite{Del16} for the $^{10}$Be($d$,$p$)$^{11}$Be case, and the effect was found to strongly depend on the deuteron incident energy as well as on the separation energy of the residual nucleus. 

Since the analysis of transfer reactions is usually performed within DWBA, ADWA and extended versions of them (such as CCBA), an important question that arises is whether these effects found in Faddeev calculations can be also described within the former approaches or are genuine to the Faddeev formulation. The clarification of this question is one of the purposes of the present work. 
For that, one must first notice that, within a coupled-channels formulation of the transfer process, one may distinguish several sources of core excitation. First, the deformation on the $n$-A potential, which gives rise to core-excited admixtures in the states of the composite nucleus $B$. This is a structure  effect, not related to the reaction mechanism, and can be properly described through spectroscopic factors obtained from a structure model including core excitation. Second, excitations of $A$ ($B$) taking place in the entrance (exit) channel, which are associated with the coupled-channels effects described above. Finally, the effect of the deformation of the proton-target potential appearing in the so-called {\it remnant term} of the DWBA or CCBA transition operator. This latter effect was already studied within an extended DWBA method \cite{Gom15} and found to be very small for the $^{10}$Be($d$,$p$)$^{11}$Be reaction.  %\sout{In a recent work \cite{Gom15} we presented an extended DWBA method which takes into account this latter effect, and found that its influence on the calculated cross section is in general small.} 
Thus, this term does not seem to be responsible for the strong dynamical effects observed in the Faddeev calculations.

In the present work, we aim at investigating the other two effects within a unified framework. An essential ingredient of the present formalism is the use of an extended CDCC method, recently revisited in Ref.~\cite{Gom16}, which provides a description of the $d$+A channel including simultaneously the effect of deuteron breakup and target excitation. This extended CDCC wave function is used within a  CCBA-like framework to calculate the $A(d,p)B$ stripping cross sections. The formalism will be applied to the 
reaction $^{10}$Be($d$,$p$)$^{11}$Be, at different deuteron energies, and the results compared with those from Ref.~\cite{Del16}.

The paper is organized as follows. In Sec.~\ref{sec:formalism} 
we outline the formal aspects of the proposed method.  
 In Sec.~\ref{sec:calc}, the formalism is 
applied to the reaction $^{10}$Be($d$,$p$)$^{11}$Be, comparing with the same observables studied in Ref.~\cite{Del16} with the Faddeev formalism, and with emphasis on the role of target excitations. In Sec.~\ref{sec:adwa}, these same observables are compared with the much simpler adiabatic approximation. In Sec.~\ref{sec:abs} we compare the three models, namely, Faddeev, CCBA and adiabatic, for the absolute transfer cross sections. 
Finally, in Sec.~\ref{sec:sum} we summarize the main results of this work.

%Then,in Sec. IV, a comparison is made between the results of the CDCC-BA formalism and the adia- batic ACC [13] one, while in Sec. V a brief com-ment on the cross sections obtained through Fad-deev, CDCC-BA and ACC formalisms is made.

\begin{figure}[tb]
\begin{center}
% {\def\svgwidth{0.9\columnwidth}{\input{zrcoor.pdf_tex}} \par}
 {\centering \resizebox*{0.85\columnwidth}{!}{\includegraphics{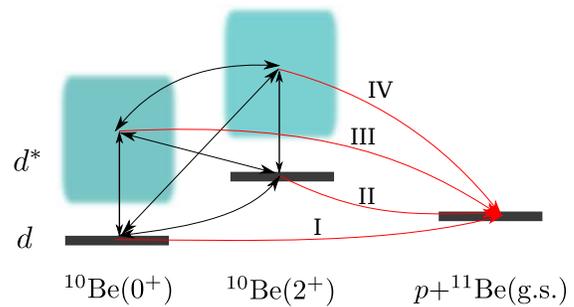}} \par}
\caption{\label{fig:scheme}(Color online) Pictorial representation of the $^{10}$Be(d,p)$^{11}$Be reaction, highlighting the different paths considered in this work. The initial $d+$\nuc{10}{Be} system includes breakup of the deuteron and excitation of \nuc{10}{Be} coupled to all orders. The transfer channel is then coupled in first order to these channels, and four paths are distinguished: I (elastic transfer), II (inelastic transfer), III (breakup transfer) and IV (inelastic breakup transfer) (See text for details.)}
\end{center}
\end{figure}

%--------------------------------------------------------------------------
\section{Theoretical framework \label{sec:formalism}}
%----------------------------------------------------------------------------

Using the post-form representation, the transition matrix for the process $A(d,p)B$ can be written as
\begin{equation}
\label{eq:T}
{\cal T}_{dp}= \langle \chi^{(-)}_p \Phi_B | V_{pn} + U_{pA}-U_{pB} | \Psi^{(+)}_d \rangle  ,
\end{equation}
where $V_{pn}$, $U_{pA}$ are the proton-neutron and proton-target interactions, $U_{pB}$ is an auxiliary (and, in principle, arbitrary) potential for the $p$-$B$ system, $\Phi_B$  is the internal wave function of the residual nucleus $B$, $\chi^{(-)}_p$ is the wavefunction for the outgoing proton, distorted by potential $U_{pB}$, and $\Psi^{(+)}_d$ is the total wave function corresponding  to an incident deuteron beam of kinetic energy $E_d$ and binding energy $\varepsilon_d$, and satisfies 
\begin{equation}
\label{eq:Sch}
[E + i \epsilon - H] \Psi^{(+)}_d(\vecr,\vecR,\xi)= i \epsilon \phi_d(\vecr) e^{i \vecK \cdot \vecR}
\end{equation}
with $E=E_d - \varepsilon_d$, $\epsilon \rightarrow 0^+$, $\xi$ the internal coordinates of $A$ and $H$ the effective Hamiltonian
\begin{equation}
\label{eq:Heff}
%H_\mathrm{proj}(\br)
H= \hat{T}_r +V_{pn} + H_A(\xi) + \hat{T}_{R}  + U_{pA}(\vecr_{pA},\xi) + U_{nA}(\vecr_{nA},\xi) ,
\end{equation}
with $\hat{T}_r$ and $\hat{T}_R$ the kinetic energy operators associated with the proton-neutron and deuteron-target relative motions. 

Ignoring antisymmetrization for clarity of the notation, the wave function $\Phi_B$ for a total angular momentum $J$ and projection $M$ can be expanded in $A$ states using the usual parentage decomposition
\begin{equation}
\label{eq:AB}
\Phi^{JM}_B(\vecr_{nA},\xi)= \sum_{I,l,j} [ \phi_{l j }(\vecr_{nA}) \otimes \Phi^I_A(\xi)]_{JM} ,
% {\rm ^{11}Be(g.s.)}\rangle= \alpha |{\rm ^{10}Be(g.s.} \otimes 2s_{1/2} \rangle +..  
\end{equation}
where $I$ is the spin of $A$, $l$ and $j$ the orbital and total ($\vec{j}=\vec{l}+s$) angular momentum of the valence particle and  $\phi_{l j }(\vecr_{nA})$   is  a function describing the neutron-core relative motion. The normalization
\begin{equation}
\label{eq:SF}
{\cal S}_{lj} = \int |\phi_{l j }(\vecr_{nA})|^2 d\vecr_{nA}
\end{equation}
can be regarded as a spectroscopic factor for the configuration $\{l,j\}$.\footnote{Strictly, the functions $\phi_{l j}$ and the spectroscopic factors ${\cal S}_{lj}$ depend in general on the core state $I$ but, in the present case, this quantum number can be readily inferred from the $l,j$ values, so it is omitted for brevity.}

We note that our starting point differs from previous few-body approaches \cite{Ron70,Ron74,Mor09,Kee04} in which possible excitations of the target nucleus are only effectively taken into account by means of the effective potentials $U_{pA}$ and $U_{nA}$. In the present work, the effective Hamiltonian (\ref{eq:Heff}) retains the dependence of the target degrees of freedom and hence its many-body nature.  
%with boundary conditions corresponding to an incoming deuteron on the target in the ground state, and outgoing waves in all possible channels. This o reduce the problem to a tractable form, this function is t We describe the process $A(d,p)B$ with the effective Hamiltonian

To solve (\ref{eq:Sch}) we use the method recently followed in our previous work \cite{Gom16}, in which $\Psi^{(+)}_d$ is approximated by a CDCC wave function and, as such, expanded in a basis of projectile (deuteron) and target ($A$) states. To make the calculation numerically tractable, the deuteron continuum is truncated in energy and angular momentum and further discretized in energy bins and the target states are restricted to the ground state and the first excited state (assumed to have spins $I=0$ and $I=2$, respectively). For clarity, we use an abbreviated notation omitting angular momentum couplings. We refer the reader to Ref.~\cite{Gom16} for further details.  This expansion reads
\begin{align}
\label{eq:Phi}
\Psi^{(+)}_d(\vecr,\vecR,\xi) & = \phi_d(\vecr) \Phi^{0}_{A} (\xi) \chi^{(+)}_{d,0}(\vecR)    \nonumber \\
             & + \phi_d(\vecr) \Phi^{2}_{A} (\xi) \chi^{(+)}_{d,2}(\vecR) \nonumber \\
             & + \sum_{i} \phi^{i}_{pn}(\vecr) \Phi^{0}_{A} (\xi) \chi^{(+)}_{i,0}(\vecR) \nonumber \\
             & + \sum_{i} \phi^{i}_{pn}(\vecr) \Phi^{2}_{A} (\xi) \chi^{(+)}_{i,2}(\vecR)
\end{align}
where $ \{\phi_d, \phi^{i}_{pn} \}$ denote the deuteron ground state and (discretized) continuum states,  and  $\{\chi^{(+)}_{i,I}(\vecR) \}$ the functions describing the projectile-target relative motion with the target in either its ground state ($I=0$) or in the excited state ($I=2$), respectively. 
 Therefore, the first two terms of (\ref{eq:Phi}) describe, respectively, elastic and inelastic scattering with the deuteron remaining in its ground state. The third  and fourth terms describe deuteron breakup with respect to the target in its ground state or first excited state, respectively.  When inserted into Eq.~(\ref{eq:T}) this gives rise also to four terms, 
\begin{equation}
\label{eq:Tdecomp}
{\cal T}_{dp}= {\cal T}^\mathrm{el}_{dp} + {\cal T}^\mathrm{inel}_{dp} +{\cal T}^\mathrm{elbu}_{dp} + {\cal T}^\mathrm{inbu}_{dp} ,
\end{equation}
which may be interpreted as (I) {\it elastic transfer}, i.e.,  direct transfer from the deuteron ground state leaving the target in its ground state, (II) {\it inelastic transfer}, i.e., target excitation followed by one-neutron transfer, (III) {\it elastic breakup transfer}, i.e.,  deuteron breakup followed by transfer, leaving the target in the ground state and (IV)  {\it inelastic breakup transfer}, i.e., deuteron breakup, accompanied by target excitation, and followed by neutron transfer.

 Note that these four terms are to be added coherently, giving rise to interference effects, as discussed below.  Note also that, in DWBA, only  the first term ({\it elastic transfer}) is explicitly  taken into account. One of our goals here is to study the contribution of the other terms. The resultant transition amplitude obtained by inserting the CDCC expansion into Eq.~(\ref{eq:T}) will be referred to as CDCC-BA approximation. We note that a similar approach has been recently used to study the $(d,p)$ reaction on \nuc{54}{Cr} \citep{Chau11}.

%------------------------------------------
\section{Results \label{sec:calc}}
%------------------------------------------

In this section we apply the aforementioned formalism to the $^{10}$Be$(d,p)^{11}$Be reaction at various deuteron energies. For our calculations, we follow a prescription as close as possible to that of \citep{Del16}, using CH89 parametrization \citep{Varner199157} for the potentials $V_{n^{10}\mathrm{Be}}$ and $V_{p^{10}\mathrm{Be}}$, evaluated at a neutron energy of $E_d/2$ and a proton energy corresponding to the exit energy of the proton. 
%The binding potential for \nuc{11}{Be} is chosen as in \citep{Del16} for a  binding energy of \nuc{11}{Be} of $0.5$~MeV, considering only the first excited state $2^+$ for \nuc{10}{Be} with an energy of $3.368$~MeV and a deformation length of $\delta_2=1.664$~fm \citep{Nunes199643}. It must be noted that in \citep{Del16} the potential used to generate the bound state for \nuc{11}{Be} must be consistent with that used in the scattering states. Thus, the potential for $n$-\nuc{10}{Be} waves which support bound states was taken as the binding potential, while the others followed the prescription described above. However, in the calculations presented here, all partial waves for $n$-\nuc{10}{Be} are treated with the CH89 parametrization.

In principle, the potential $U_{pA}$ appearing in the transition operator of the transfer amplitude 
(\ref{eq:T}) should retain its dependence on the $\xi$ coordinates and would be non-central, thereby permitting transitions between different core ($A$) states, not present in the standard DWBA and CCBA implementations. However, in our previous work \cite{Gom15} we studied the effect of this {\it core excitation} mechanism for the $^{10}$Be$(d,p)^{11}$Be   reaction and it was found to be very small,  so we will omit it in the calculations presented in this work, to avoid the formal and numerical complications that it introduces.

The interaction $V_{pn}$ has been chosen of a Gaussian shape as in \citep{PhysRevC.85.054621}, while in \citep{Del16}, CD Bonn potential \citep{PhysRevC.63.024001} is used. However, in \citep{Del16} it is mentioned that the use of the Gaussian potential leads to differences within $2\%$, so we expect this difference in potentials to be of little relevance. The interaction $V_{p^{11}\mathrm{Be}}$ is obtained from the CH89 parametrization at the exit energy of the proton, while in the Faddeev/AGS calculation all the dynamics is generated from the two-body interactions. 
%it is obtained through $V_{pn}$ and $V_{p^{10}\mathrm{Be}}$. 
Finally, we must indicate that in \citep{Del16} a subtraction technique is applied to the $n/p$+$^{10}$Be interactions  to preserve the elastic nucleon-core cross section \citep{Del15}. Since this technique results in non-local potentials, we have not applied it to our calculations, where non-local potentials cannot be used.

% AMM
The structure of the $^{11}$Be nucleus is treated within the particle-rotor model of Ref.~\citep{Nunes199643}, which assumes a deformation length of $\delta_2=1.664$~fm for  $^{10}$Be and includes the ground state ($0^+$) and first excited state ($2^+$) of this nucleus. The potential parameters are adjusted to give the experimental neutron separation energy ($S_n=0.5$~MeV). In this model, the $^{11}$Be ground state wave function can be expressed as  in Eq.~(\ref{eq:AB})
%consists of a superposition of $\ell=0$ and $\ell=0$ configurations
%
\begin{align}
\label{eq:be11gs}
\Phi^{\half M}_{B}(\vecr_{nA},\xi) & = [\phi_{0,\half}(\vecr_{nA}) \otimes \Phi^0_A(\xi) ]_{\half M} \nonumber \\ 
 & +   [\phi_{2,\fhalf}(\vecr_{nA}) \otimes \Phi^2_A(\xi) ]_{\half M} \nonumber \\ 
 & +   [\phi_{2,\thalf}(\vecr_{nA}) \otimes \Phi^2_A(\xi) ]_{\half M}, 
% {\rm ^{11}Be(g.s.)}\rangle= \alpha |{\rm ^{10}Be(g.s.} \otimes 2s_{1/2} \rangle +..  
\end{align}
with $B$ and $A$ denoting the $^{11}$Be and $^{10}$Be nuclei and $\phi_{l j}(\vecr_{nA})$  the overlap functions between them,  with weights ${\cal S}_{lj}$=0.846, 0.130, 0.023 for the $s_{1/2}$, $d_{5/2}$ and $d_{3/2}$ components, respectively. 

% 
%\sout{It must be noted that in Ref.~\citep{Del16} the potential used to generate the bound state for \nuc{11}{Be} must be consistent with that used in the scattering states. Thus, the potential for $n$-\nuc{10}{Be} waves which support bound states was taken as the binding potential, while the others followed the prescription described above. However, in the calculations presented here, all partial waves for $n$-\nuc{10}{Be} are treated with the CH89 parametrization.}
It must be noted that, in our formalism, the potential for $n$-\nuc{10}{Be} is taken differently for the entrance and exit channels. In the entrance channel, it is represented by a  complex optical potential (CH89) whereas in the exit channel it is represented by a real potential used to generate the $^{11}$Be bound state. By contrast, in the  Faddeev formalism of Ref.~\citep{Del16}, there is no separation between entrance and exit Hamiltonians but this interaction is taken to be $l$-dependent; real,  for the partial waves supporting the $^{11}$Be bound states, and complex (CH89) for the other waves. 

To compare with the results of Ref.~\citep{Del16} we have computed the ratio $R_x$ defined as in that reference as
\begin{equation}
\label{eq:rx}
R_x=\dfrac{1}{{\cal S}_F}\dfrac{\left(\dfrac{d\sigma}{d\Omega}\right)^{\rm peak}_{\rm def}}{\left(\dfrac{d\sigma}{d\Omega}\right)^{\rm peak}_{\text{no def}}},
\end{equation}
where ${\cal S}_F$ is the spectroscopic factor associated to the \nuc{10}{Be}($0^+$) component in the structure model for \nuc{11}{Be} including deformation (in our case, \ ${\cal S}_F = {\cal S}_{0,\half}$), while $\left({d\sigma}/{d\Omega}\right)^{\rm peak}$ is the differential cross section for  $^{10}$Be$(d,p)^{11}$Be transfer at its peak for the calculations including deformation (as indicated in the previous section) and excluding it, when corresponding. In the absence of dynamical effects, $R_x$ is expected to be 1.

The calculations have been performed discretizing the deuteron continuum through a binning procedure \cite{Aus87} up to the center of mass energy of the system, using 4-6 bins, evenly spaced in momentum space. Breakup states with $l=0,2$ have been considered, since the inclusion of $l=1,3$ led only to a modest modification of the cross sections in test calculations at $E_d=20$~MeV and 60~MeV.
 
It has been found that the $R_x$ factors are rather insensitive to the discretization used, as long as the same one is used in both the calculation including deformation and that excluding it. We also found that the $R_x$ factors achieved convergence within a few percent using broader meshes and lower maximum deuteron excitation energies than those required for the actual cross sections to converge.

\begin{figure}
\includegraphics[width=0.45\textwidth]{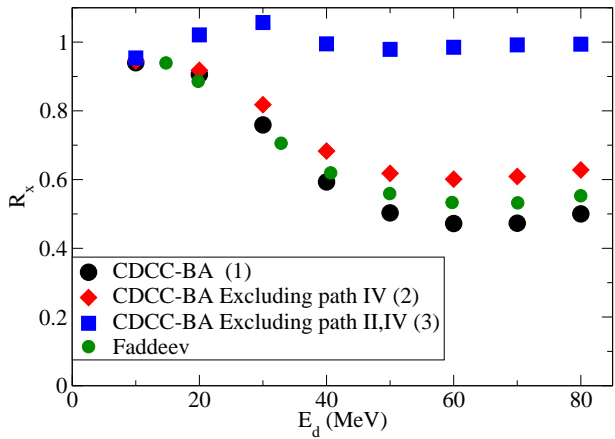}
 \caption{\label{fig:rx} (Colour online)  $R_x$ factors (see text) for different deuteron energies. The green circles correspond to the results from Faddeev/AGS calculations \citep{Del16}. In black circles the factors for the full CDCC-BA calculations are presented, while the red diamonds correspond to calculations where ${\cal T}^{inbu}_{dp}$ (path IV from Fig.~\ref{fig:scheme}) is blocked and the blue squares correspond to calculations where ${\cal T}^{inel}_{dp}$ and ${\cal T}^{inbu}_{dp}$ are blocked (II and IV from Fig.~\ref{fig:scheme}). See text for details.} 
\end{figure}

In Fig.~\ref{fig:rx}, the factors $R_x$ are presented as a function of the deuteron laboratory energy. The green circles correspond to the values obtained in \citep{Del16} from the Faddeev calculations. Three CDCC-BA calculations are presented. As indicated in the previous section, in all of them all $d+$\nuc{10}{Be} states are coupled to all orders (a scheme of the process is depicted in Fig.~\ref{fig:scheme}), and the difference between them lies on the couplings included for the transfer process itself. The black circles (calculation 1) correspond to the full calculation, which includes the four terms  in the transition matrix  (\ref{eq:Tdecomp}). This calculation reproduces rather well the Faddeev results, particularly at the lower deuteron energies, having a slight underestimation at the larger deuteron energies. 

As stated in the introduction, one of goals of this work is to shed light on the origin of the deviation from unity of the $R_x$ factor. For that, we have performed further calculations, in which some of the transfer couplings are selectively omitted. The results of these calculations are also shown in Fig.~\ref{fig:rx}. The red diamonds (calculation 2) correspond to the calculation excluding transfer from states where both deuteron and \nuc{10}{Be} are excited (path IV in Fig.~\ref{fig:scheme}), so that the transition matrix for the transfer process is:
\begin{equation}
\label{eq:T2}
{\cal T}^{(2)}_{dp}= {\cal T}^\mathrm{el}_{dp} + {\cal T}^\mathrm{inel}_{dp} +{\cal T}^\mathrm{elbu}_{dp}.
\end{equation}
This calculation gives similar values of $R_x$ at the lower incident energies, and a moderate increase of $R_x$ at the largest energies when compared to the full calculation, although a significant reduction with respect to unity persists. This clearly indicates that the transfer via the $p$+$n$+$^{10}$Be$^*$ channels, while not completely negligible, is not responsible for the behaviour of the $R_x$ factor. 

Finally, the blue squares (calculation 3) correspond to a calculation in which all states with \nuc{10}{Be} in its excited state are excluded from ``feeding'' the transfer channel (i.e.\ excluding paths II and IV in Fig.~\ref{fig:scheme}), leading to the transition matrix:
\begin{equation}
\label{eq:T3}
{\cal T}^{(3)}_{dp}= {\cal T}^\mathrm{el}_{dp} + {\cal T}^\mathrm{elbu}_{dp}.
\end{equation}
This calculation deviates significantly from the previous ones, giving $R_x$ factors close to 1, specially at higher energies. This may seem trivial, since blocking transfer from states with \nuc{10}{Be} in its excited state would be expected to give the same cross section than the calculation without deformation. However it must be noted that $\Psi^{(+)}_d$ is different in both calculations, since it includes coupling to the excited state of \nuc{10}{Be} in the calculation with deformation but it excludes it in the calculation without deformation. Therefore, the factor $R_x$ in calculation 3 would be expected to be senstitive to the modifications in $\Psi^{(+)}_d$ due to deformation, at least in the region that is relevant for transfer. The fact that it is close to 1 suggests that the effects of deformation on it are small in this region. This interpretation seems to be consistent with the fact that $R_x$ for this  calculation  get closer to 1 at higher energy, where coupling effects in $\Psi^{(+)}_d$ are expected to be smaller.
 
%%%%%%%%%%%%%%%%%%%%%%%%%%%%%%%%%%%%%%%%%%%%%%%%%%%%%%%%%%%%%%%%%%%%%%%%%%%%%%%%%%%%%%%%%%%%%%%
\begin{comment} % AMM
so that the transition matrix is that of Eq.~(\ref{eq:Tdecomp}). The difference among the calculations lies on the coupling to the $p+$\nuc{11}{Be} partition.  where all $d+$\nuc{10}{Be} channels are coupled in first order to the transfer channel. This includes the four terms in Eq.~(\ref{eq:Tdecomp}). Meanwhile, the red diamonds (Calculation 2) correspond to the calculation excluding the transfer from the 
%
Finally, 

We find that calculations 1 and 2 reproduce well the tendency obtained from Faddeev calculations, although none of them gives the same absolute values, with the full calculation seeming to give a slightly better agreement. As for calculation 3, it 
\end{comment}

\begin{figure}
\includegraphics[width=0.4\textwidth]{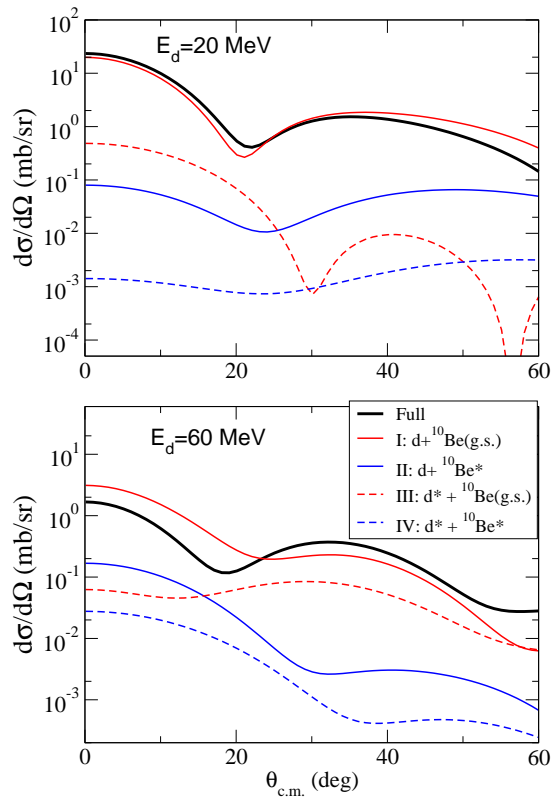}
 \caption{\label{fig:tr_desc} (Colour online) Differential transfer cross section for $E_d=$20~MeV (top) and 60~MeV (bottom). All results correspond to full CDCC calculations for the elastic channel, while the transfer channel has been calculated through Born approximation from all paths, and paths I, II, III and IV (see Fig.~\ref{fig:scheme}) for the black solid, red solid, blue solid, red dashed and blue dashed lines respectively. (See text)} 
\end{figure}

In order to further clarify the effects of inelastic and breakup channels on transfer, we show in Fig.~\ref{fig:tr_desc} the angular distribution of the transfer cross section at two deuteron energies, $E_d=20$ MeV (top) and 60 MeV (bottom). In each panel, we show the full calculation (thick solid black line) along with the calculations keeping only one of the terms in Eq.~(\ref{eq:Tdecomp}). The red solid line corresponds to ${\cal T}^{el}_{dp}$ (path I in Fig.~\ref{fig:scheme}), the blue solid  only to ${\cal T}^{inel}_{dp}$ (II), the red dashed only to ${\cal T}^{elbu}_{dp}$ (III) and the blue dashed only to ${\cal T}^{inbu}_{dp}$ (IV). As before, for the $d$+\nuc{10}{Be} partition all channels are coupled to all orders, while the transfer channel is coupled to first order. 
  It can be seen that for both energies the main contributor is  the elastic transfer ${\cal T}^{el}_{dp}$, while, at small angles,  for $E_d=20$ MeV  the second main contributor  is the breakup of the deuteron, ${\cal T}^{elbu}_{dp}$, whereas at 60 MeV it is the excitation of \nuc{10}{Be}, ${\cal T}^{inel}_{dp}$ . This seems to agree with the fact that $R_x$ is smaller at higher energies, since a greater effect of \nuc{10}{Be} excitation should lead to a reduced $R_x$. 

\begin{figure}
\includegraphics[width=0.4\textwidth]{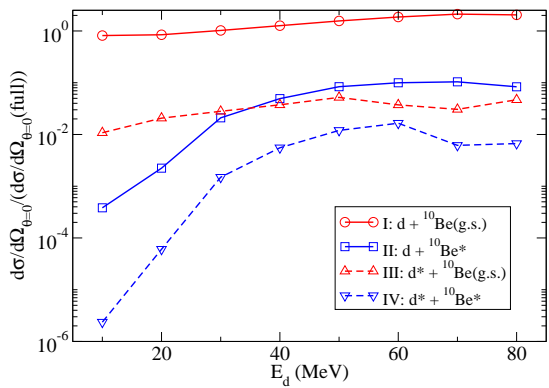}
 \caption{\label{fig:th0_vs_E} (Colour online) Ratio between cross sections at the peak  as a function of the deuteron incident energy. For each energy the ratio is between the calculation where one of the transfer paths is selected (as in Fig.~\ref{fig:tr_desc}) and the full calculation. As can be seen, path II, corresponding to the inelastic excitation of \nuc{10}{Be}, gains importance with increasing energy so that at the higher energies, where the $R_x$ factors are smaller, it is of the same magnitude as path III, corresponding to deuteron breakup.} 
\end{figure}

In order to test this relation between $R_x$ and the importance of the inelastic path, we plot in Fig.~\ref{fig:th0_vs_E}, as a function of incident energy, the ratio of the cross section at the peak for the calculation where one of the transfer paths was selected divided by the cross section at the peak for the full calculation, containing all transfer paths. As can be seen in the figure, path II, corresponding to ${\cal T}^{inel}_{dp}$ gains relevance as energy increases, and we note that the energies where its importance is higher are those with lower $R_x$ so a relation can be established, albeit qualitative, between the importance of the inelastic path and the value of $R_x$.

At both energies shown in Fig.~\ref{fig:tr_desc}, simultaneous excitation (i.e. deuteron breakup concurrent with target excitation), ${\cal T}^{inbu}_{dp}$, is the least important component. It must be noted nevertheless that, due to the interference of the different transfer paths, the relevance of each channel cannot be directly inferred from the cross section shown in Fig.~\ref{fig:tr_desc}. If this were the case, the small cross section of the concurrent excitation ${\cal T}^{inbu}_{dp}$ would lead to a $R_x$ factor for calculation 2 much closer to that of the full calculation than what is actually obtained. Therefore we may conclude that interference between the different transfer paths is of relevance for these reduction factors.

\begin{figure}
\includegraphics[width=0.4\textwidth]{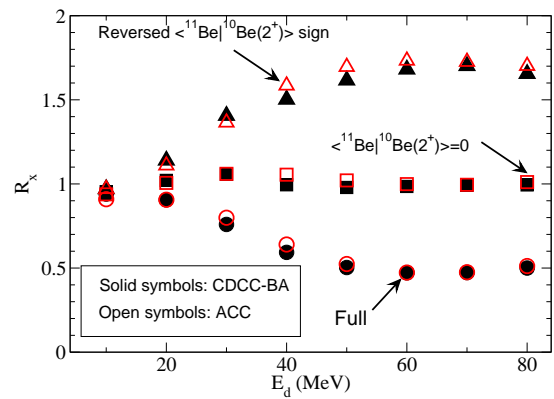}
 \caption{\label{fig:rx_adwa} (Colour online) $R_x$ ratio, as defined in Eq.~(\ref{eq:rx}). The black solid symbols correspond to calculations using a CDCC-BA formalism while the red empty ones are obtained from a coupled-channel adiabatic calculation (ACC). The circles, squares and triangles refer to the full calculations, the calculations where the transfer from \nuc{10}{Be}($2^+$) has been blocked, and the calculations where the sign of the overlaps involving \nuc{10}{Be}($2^+$) has been reversed, respectively (see text). }
\end{figure}

The fact that calculation 3 gives reduction factors $R_x$ so close to 1, suggests  that the main responsible for the reduction in the full calculation  (calculation 1) is the interference between the transfer via \nuc{10}{Be}($0^+$), that is, ${\cal T}^{el}_{dp}$ and ${\cal T}^{elbu}_{dp}$, and that via the \nuc{10}{Be}($2^+$) components, ${\cal T}^{inel}_{dp}$ and ${\cal T}^{inbu}_{dp}$. To test this conclusion, we have reversed the sign of the \nuc{10}{Be}($2^+$) components of the \nuc{11}{Be} g.s.\ wave function, Eq.~(\ref{eq:be11gs}), and computed the reduction factors once again. The result of this calculation is shown in Fig.~\ref{fig:rx_adwa} in black solid triangles, along with the result of the original full calculation, in black solid circles and that of calculation 3, which corresponds to neglecting the \nuc{10}{Be}($2^+$) components, in black solid squares. We find that reversing the sign of the \nuc{10}{Be}($2^+$) components gives $R_x$ greater than 1 which show a reversed tendency from that obtained for the original calculation. This confirms that the origin of the reduction of the cross section lies in the destructive interference of the transfer to components of \nuc{11}{Be} with \nuc{10}{Be} in its ground or excited states, and that their relative sign is critical to describe its behaviour.

The fact that both Faddeev and CDCC-BA calculations give similar results for $R_x$ suggests that these factors are not very sensitive to the formalism used for the description of the reaction. Therefore, in the following section we apply a simpler adiabatic formalism to the same reaction and analyse the obtained reduction factors $R_x$.

%-----------------------------------------------
\section{Comparison with the adiabatic model \label{sec:adwa}}
%-----------------------------------------------
The results of the preceding section indicate that the departure of the calculations with deformation from those without it is mainly due to the inelastic transfer mechanism, i.e., the two-step transfer taking place via the target excited state (path II in Fig.~\ref{fig:scheme}). This does not mean that breakup channels are not important, since they are essential to give a correct description of the elastic scattering. 
This result suggests that, insofar as the transfer cross section is concerned, one might resort to a simpler procedure in which target excitation is treated explicitly, whereas breakup channels are accounted for effectively. This can be suitably done within the adiabatic approximation. Although the original formulation of this method \cite{Ron70,Har70,Ron74}, usually referred to as adiabatic distorted wave approximation (ADWA),   does not include explicitly the effect of target excitation, the model can be generalized to accommodate this effect. 
%This has been done in the finite-range approximation \cite{Del05} but we resort here to the simpler zero-range approximation, of Johnson and Soper (JS) \cite{Ron70}. 
 As in the extended CDCC method, the extended adiabatic  model can be derived 
from the generalized Hamiltonian (\ref{eq:Heff}) using  the same procedure of Refs.~\cite{Ron74,Lai93}. 
The three-body wave function is then expanded  in a Weinberg basis \cite{Ron74,Lai93} whose leading term gives rise to the following approximate three-body wave function:
\begin{equation}
\label{eq:PhiAD}
\Psi^\mathrm{AD}_d(\vecr,\vecR,\xi)= \phi_d(\vecr)  [ \Phi^{0}_{A}(\xi) \chi^\mathrm{AD}_{0}(\vecR) 
+ \Phi^{2}_{A}(\xi) \chi^\mathrm{AD}_{2}(\vecR)  ]  , 
\end{equation}
where $\{ \chi^\mathrm{AD}_{0}(\vecR), \chi^\mathrm{AD}_{2}(\vecR)\}$  are solutions of a set of coupled equations with a deformed adiabatic deuteron-target  potential. This has been done in the finite-range approximation \cite{Del05} but we resort here to the simpler zero-range approximation. In this limit, the deformed adiabatic potential results
%Although this has been done in finite-range \cite{Del05},  we resort here to the simpler zero-range version.  In this limit, the deformed adiabatic potential results
\begin{equation}
\label{eq:Vadwa}
U^\mathrm{AD}(\vecR,\xi)= V^\mathrm{AD}_0(\vecR) - \delta_\lambda Y_{\lambda,0}(\xi) Y_{\lambda,0}(\vecR) \frac{d V^\mathrm{AD}_0(\vecR)}{dR} ,
\end{equation}
with
\begin{equation}
\label{eq:V0adwa}
 V^\mathrm{AD}_0(\vecR)  = U^0_{pA}(\vec{R},\xi) + U^0_{nA}(\vec{R},\xi) ,
\end{equation}
which is nothing but the generalization of the Johnson and Soper prescription \cite{Ron70} for  deformed nucleon-nucleus potentials and where $U^0_{p(n)A}$ indicate the central part of the proton (neutron)-target potential. 
%The first term in (\ref{eq:Vadwa}) is the Johson-Tandy adiabatic finite-range potential, whereas the second term arises from the deformation of the underlying nucleon-target potentials. 
The method has been referred to before as adiabatic coupled channels (ACC) method \cite{Del05}.

We have computed the quantity $R_x$, starting from the definition (\ref{eq:rx}),  but using now for the numerator and denominator  the ACC and ADWA methods, respectively. The results are shown in Fig.~\ref{fig:rx_adwa} (open circles).
% For comparison, we include also the CDCC-BA results of Fig.~\ref{fig:rx} (solid circles).  
The agreement with the CDCC-BA results (solid circles) is remarkably good, confirming that  core excitation effects are well accounted for by the much simpler adiabatic model. To further delineate  the role played by the target excited state in the stripping process, we include also the calculations in which the transfer via the excited state (path II) is omitted and those in which the sign of the $l=2$ components [$\phi_{2,\fhalf}$ and $\phi_{2,\thalf}$  in Eq.~(\ref{eq:be11gs})]  are reversed (open squares and open triangles, respectively). Again, the agreement with the CDCC-BA results (solid symbols) is very good.   
%This gives rise to the open circles which, as in the CDCC-BA calculation, is close to unity.  

%-----------------------------------------------
\section{Comparison of absolute cross sections \label{sec:abs}}
%-----------------------------------------------

\begin{figure}
\includegraphics[width=0.4\textwidth]{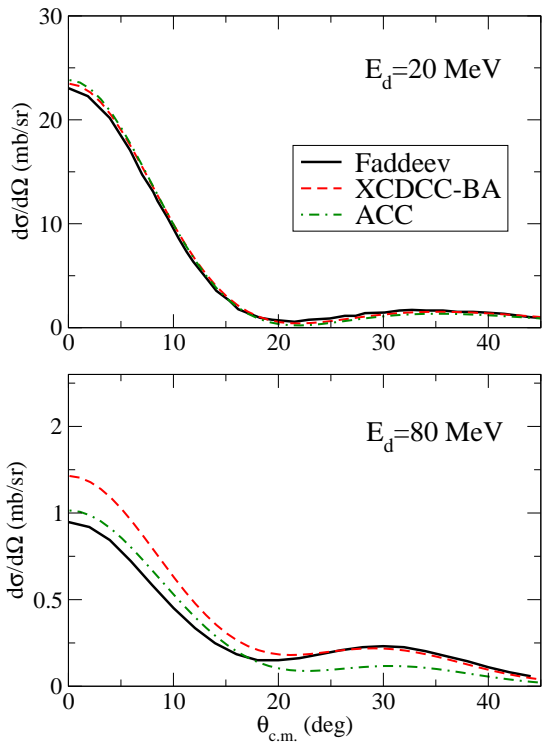}
 \caption{\label{fig:tr_dec} (Colour online) Differential cross section angular distributions at $E_d$=20~MeV (top) and $E_d$=80~MeV (bottom) for the  $^{10}$Be$(d,p)^{11}$Be calculated with the Faddeev/AGS \citep{Del16}, CDCC-BA and ADWA formalisms.} 
\end{figure}

In the previous sections, we have found that both the CDCC-BA and ACC methods successfully describe the behaviour of the $R_x$ ratio as a function of the deuteron energy, as compared to the more sophisticated Faddeev/AGS calculations. Despite these encouraging results, previous benchmark calculations have evidenced limitations of the  CDCC and adiabatic approaches in the reproduction of the Faddeev absolute cross sections \cite{Del07,Upa12}. In particular, for transfer reactions, it was found that the agreement tended to deteriorate with increasing incident energies \cite{Upa12}. We have compared these three reaction models in presence of core excitations. This comparison is shown in Fig.~\ref{fig:tr_dec} for two deuteron energies, $E_d=20$~MeV (upper panel) and $E_d=80$~MeV (bottom panel). At $E_d=20$~MeV, both the CDCC-BA and ACC formalisms reproduce very well the Faddeev result, with the CDCC-BA model providing a somewhat better agreement close to zero degrees. At $E_d=80$~MeV, the agreement is deteriorated, with both the CDCC-BA and ACC overestimating the small-angle cross sections. Further, the ACC curve underpredicts the maximum at 30$^\circ$. Interestingly, these results are qualitatively similar to those found in \cite{Upa12} for the undeformed case. In particular, the agreement between the formalisms at smaller deuteron energies is to be highlighted. In \cite{Oga16} it was found that a similar disagreement between CDCC and Faddeev calculations for breakup observables was substantially improved when including closed channels in the CDCC calculation. Due to computational difficulties, a similar study was not possible in this work. However, since our main quantities of interest, the reduction factors $R_x$, give a good agreement for the three considered formalisms and show reasonable convergence in the calculations performed, we consider these considerations to be beyond the scope of this work.

%------------------------------------------------
\section{Summary and conclusions \label{sec:sum}} 
%-------------------------------------------------
To summarize, we have studied the role of target excitations in $A(d,p)B$ reactions, taking as a test case the $^{10}$Be($d$,$p$)$^{11}$Be reaction. For that, we have considered two different reaction formalisms which incorporate the effect of deuteron breakup and target excitation. 

The first method (CDCC-BA), which has been developed in this work, describes the $d+A$ system using a generalized CDCC wave function, which treats deuteron breakup and target excitation to all orders and  transfer in Born approximation. The model includes, in addition to the direct transfer coming directly from the projectile and target ground states,  the multi-step transfer through the deuteron continuum states as well as from the target excited state, so it permits the study of the role played by each of these paths in the calculated stripping cross sections. The second method (ACC) is a coupled-channel version of the zero-range adiabatic model of Johnson and Soper \cite{Ron70}. It considers also explicitly the transfer via the target ground and excited states, but it treats the effect of breakup only effectively, within an adiabatic approximation.

Following a previous work \cite{Del16}, we have studied the ratio ($R_x$) of the calculated transfer cross sections evaluated with and without deformation  at the transfer peak. We have found that both the CDCC-BA and ACC methods are able to reproduce very well this ratio and its energy dependence, when compared to Faddeev/AGS calculations. The deviation of $R_x$ from unity is found to stem mostly from the interference of the elastic and inelastic transfer amplitudes. Moreover, this ratio is weakly affected by the transfer via the deuteron breakup states, so that their effect seems to be rather well simulated by the adiabatic three-body wave function. At larger angles and for high incident energies, the effect of breakup channels becomes  more significant.

The comparison of the absolute cross sections is qualitatively similar to that previously  observed in previous benchmark calculations without core excitation, namely,  at small deuteron energies, both the CDCC-BA and ACC methods reproduce rather well the Faddeev results, but the agreement worsens  at incident energies of several tens of MeV and above.  Theoretical works oriented to better understand the origin of these limitations and envisage possible improvements, both with and without core excitation effects, are of major relevance to reliably apply these methods to the analysis of transfer reactions.

%
% Acknowledgements -------------------------------------------------------------- 
% Note that in elsevier documentclass style, there is a command 
% \ack for this purpose!!!!
%
\begin{acknowledgments}
We are grateful to A.~Deltuva and F.~Delaunay useful discussions.   
This work has been partially supported by the Spanish
 Ministerio de  Econom\'ia y Competitividad and FEDER funds under project 
 FIS2014-53448-C2-1-P  and by the European Union's Horizon 2020 research and innovation program under grant agreement No.\ 654002. M.~G.-R. acknowledges a research grant from the Spanish Ministerio de Educaci\'on, Cultura y Deporte, Ref: FPU13/04109.
\end{acknowledgments}

%--------------------------------------------------------------------------------------------------

%% References


\bibliography{transfer}
\end{document}